\documentclass[12pt]{article}

\def\spacingset#1{\renewcommand{\baselinestretch}%
{#1}\small\normalsize} \spacingset{1}

\usepackage{graphicx} 
\usepackage{subfigure}

\usepackage{amsmath, amsfonts, amssymb, amsthm, amsbsy, amscd, bm, bbm, mathrsfs}
\usepackage{algorithm, algorithmic}
\usepackage{enumitem} 
\usepackage{url,hyperref}
\usepackage{cleveref}
\usepackage{subcaption}
\usepackage{wrapfig}
\usepackage{float}
\usepackage{color-edits}
\usepackage{multirow}
\usepackage{booktabs}
\usepackage{multirow}
\usepackage{soul,xcolor} 
\setstcolor{red}     
\usepackage{natbib} 
\usepackage[table,xcdraw]{xcolor}

\newtheorem{theorem}{Theorem}
\newtheorem{othertheorem}{othertheorem}[section]
\newtheorem{lemma}[othertheorem]{Lemma}

\newtheorem{claim}[othertheorem]{Claim}

\theoremstyle{definition}
\newtheorem{definition}[othertheorem]{Definition}
\newtheorem{remark}[othertheorem]{Remark}

\theoremstyle{definition}

\newcommand{\vp}{\bm{p}}
\newcommand{\vr}{\bm{r}}
\newcommand{\ve}{\bm{e}}

\renewcommand{\thefootnote}{\arabic{footnote}}

\title{
Secret-Protected Evolution for Differentially Private Synthetic Text Generation
}

\author{Tianze Wang$^{1,2}$\footnotemark[1] $\quad$ Zhaoyu Chen$^{1}$\footnotemark[1] $\quad$ Jian Du$^{1}$\footnotemark[2] $\quad$ Yingtai Xiao$^{1}$ \\ Linjun Zhang$^{2}$ $\quad$ Qiang Yan$^{1}$
\\ $^1$ TikTok $\quad^2$Department of Statistics, Rutgers University
}

\date{}

\begin{document}

\maketitle

\renewcommand{\thefootnote}{\fnsymbol{footnote}} 
\footnotetext[1]{indicates equal contributions. This work was done when Tianze Wang was an intern at TikTok.} 
\footnotetext[2]{indicates corresponding author. E-mail: jian.du@bytedance.com} 

\begin{abstract}
Text data has become extremely valuable on large language models (LLMs) and even lead to general artificial intelligence (AGI).
A lot of high-quality text in the real world is private and cannot be freely used due to privacy concerns. Therefore, differentially private (DP) synthetic text generation has been proposed, aiming to produce high-utility synthetic data while protecting sensitive information.
However, existing DP synthetic text generation imposes uniform guarantees that often overprotect non-sensitive content, resulting in substantial utility loss and computational overhead. Therefore, we propose Secret-Protected Evolution (SecPE), a novel framework that extends private evolution with secret-aware protection. 
Theoretically, we show that SecPE satisfies $(\vp, \vr)$-secret protection, constituting a relaxation of Gaussian DP that enables tighter utility–privacy trade-offs, while also substantially reducing computational complexity relative to baseline methods.
Empirically, across the OpenReview, PubMed, and Yelp benchmarks, SecPE consistently achieves lower Fréchet Inception Distance (FID) and higher downstream task accuracy than GDP-based Aug-PE baselines, while requiring less noise to attain the same level of protection. 
Our results highlight that secret-aware guarantees can unlock more practical and effective privacy-preserving synthetic text generation.

\end{abstract}

\section{Introduction}
Text data has grown immensely valuable for large language models (LLMs), enabling these models to achieve revolutionary breakthroughs in natural language understanding and generation while delivering robust performance across document-understanding tasks—including classification, contextual autocompletion, and social recommendation~\cite{10.1145/3292500.3330723, Mukherjee2020NaturalLP, 10.1007/978-3-030-85292-4_24, Harte_2023}. 
However, training and adaptation typically rely on large volumes of private user text data, raising serious privacy risks including memorization and leakage of sensitive content \cite{carlini2019secretsharerevaluatingtesting,
carlini2021extractingtrainingdatalarge, lukas2023analyzingleakagepersonallyidentifiable,
wang2024decodingtrustcomprehensiveassessmenttrustworthiness}.

To address privacy leakage, Differential Privacy (DP) \cite{10.1007/11681878_14} has become the gold standard, offering a rigorous mathematical framework for mitigating information disclosure. Therefore, synthetic text based on DP can be safely shared and used for downstream tasks. A classical approach is to train a DP generator \cite{abadi2016deep} and then sample DP synthetic data \cite{yue2023synthetic,yu2024privacypreserving,tan2025synthesizing}. Despite its conceptual simplicity, such generators are computationally intensive, require hundreds of high-quality private data to achieve strong performance, and cannot directly leverage closed-source, state-of-the-art LLMs. More recently, \emph{Private Evolution} (PE) \cite{lin2025differentiallyprivatesyntheticdata} has emerged as an alternative: rather than privately training a model, one repeatedly queries a powerful foundation model to generate candidates, evaluates them against private data via DP voting, and resamples around the winners \cite{xie2024differentiallyprivatesyntheticdata, zou2025contrastive}.

PE leverages strong off-the-shelf models and shifts the privacy cost to selection and aggregation. However, it still requires a substantial volume of private samples, and its pairwise similarity computations and iterative data processing make the pipeline highly inefficient. This inefficiency poses a critical challenge in practice and motivates the need for more scalable solutions. In addition, the reliance on canonical DP assumes that every record is equally sensitive, even though sensitive information may be sparse \cite{shi2022selectivedifferentialprivacylanguage} and vary across users and attributes (e.g., medical records vs.\ movie ratings). Furthermore, secrets may repeat across records, and a single user may contribute multiple records; under user-level DP, this further degrades utility \cite{levy2021learninguserlevelprivacy, chua2024mindprivacyunituserlevel, charles2024finetuninglargelanguagemodels}, increasing the noise required by uniform guarantees. 

\begin{figure}[t]
    \centering
    \includegraphics[width=.93\linewidth]{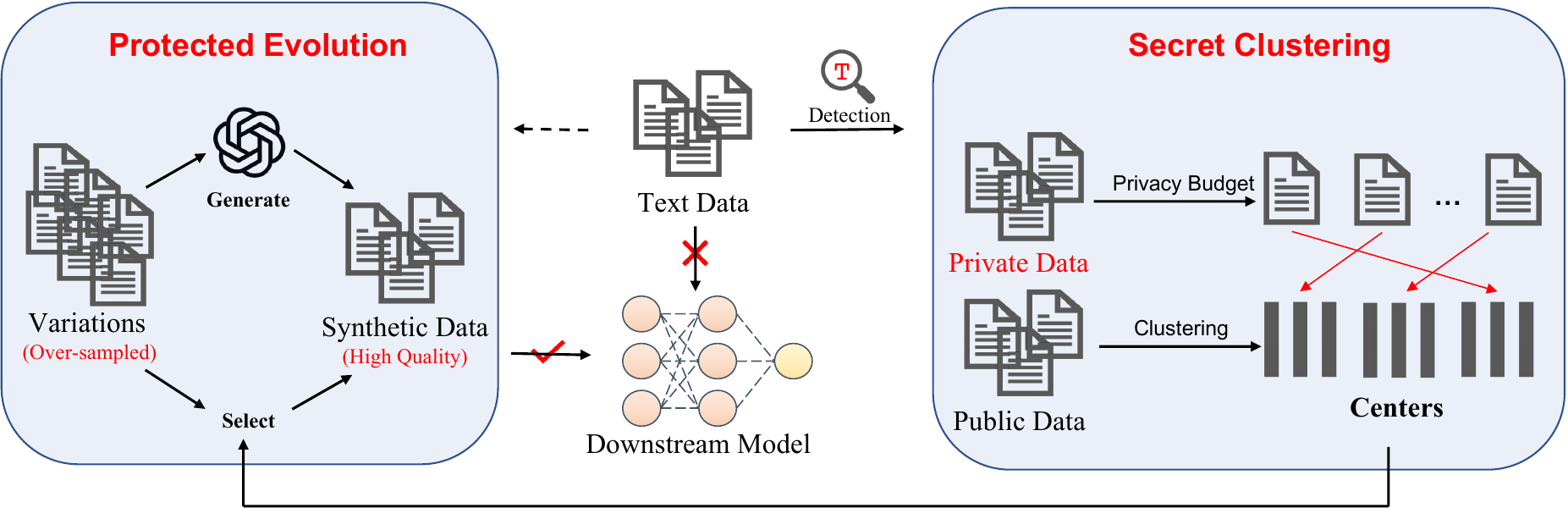}
    \caption{The overall of SecPE. The framework consists of two modules: 
    (1) Secret Clustering: clustering is applied to public data and updated with noisy private data to form representative centers for voting; 
    (2) Protected Evolution: in each iteration, candidate synthetic data consist of high-quality samples from the previous iteration together with their LLM-generated variations, and new high-quality samples are selected based on similarity to the noisy representatives.}
    \label{fig:secpe_flow}
\end{figure}

Recent work argues for \emph{secret protection}, which provides guarantees tailored to specific secrets rather than membership \cite{ganesh2025hushprotectingsecretsmodel}.
With predefined secrets, public data can be used without protection. For example, to cluster and summarize the dataset, thereby improving efficiency while reserving privacy noise for secret-related adjustments. 
In the formulation, protection is calibrated at the adversary’s prior for secrets and then directly bounding the reconstruction success probability, a similar concept as \cite{hayes2023boundingtrainingdatareconstruction}.  Conceptually, this relaxes Gaussian DP (GDP) \cite{dong2019gaussiandifferentialprivacy} by shifting from enforcing a lower bound on the \emph{entire} trade-off curve to controlling only the operative point, yielding tighter utility–privacy trade-offs while retaining meaningful reconstruction guarantees. This perspective suggests re-thinking PE around \emph{secret-aware} selection and aggregation rather than uniform DP noise.

In this paper, we introduce the \textbf{Secret-Protected Evolution (SecPE)} framework. As illustrated in Figure~\ref{fig:secpe_flow}, SecPE consists of two key components: 
(1) \emph{Secret Clustering}, which detects sensitive attributes and forms representative centers by updating public clusters with noisy private data; and 
(2) \emph{Protected Evolution}, which iteratively samples variations from high-quality synthetic data, evaluates them against the noisy representatives, and selects the best candidates. 
This design preserves the practicality of PE while shifting protection toward secrets rather than uniform DP.

Our contributions are summarized as follows: 
(1) we propose a private synthetic data generation framework that emphasizes \emph{secret protection} rather than canonical DP, thereby improving utility by reducing the noise typically required under DP; 
(2) we develop a secret-protected clustering method that substantially reduces runtime complexity compared to the PE approach, enabling scalability to larger datasets while maintaining competitive performance; 
and (3) through experiments on OpenReview, PubMed, and Yelp, we empirically demonstrate that SecPE achieves higher efficiency, lower Fréchet Inception Distance (FID)~\cite{heusel2018ganstrainedtimescaleupdate}, and better downstream accuracy than $\mu$-GDP--based PE baselines under the same reconstruction guarantees.

\section{Related Work}
\paragraph{Differential Privacy (DP).} 
We begin by reviewing $(\epsilon,\delta)$-DP and Gaussian Differential Privacy (GDP), the latter providing a clean bridge to secret-protection analysis.
\begin{definition}[$(\epsilon,\delta)$-DP] A randomized algorithm $\mathcal{M}$ is $(\epsilon, \delta)$-differentially private if, for any two neighboring datasets $D$ and $D^\prime$ that differ in exactly one datapoint, and for all $\mathcal{S} \subset \text{Range}(\mathcal{M})$, it holds that $\Pr [\mathcal{M}(D) \in \mathcal{S}] \leq e^\epsilon \Pr [\mathcal{M}(D^\prime) \in \mathcal{S}] + \delta.$
\end{definition}

Within the DP framework, an adversary aims to decide whether a specific record is present in a dataset. This naturally leads to a binary hypothesis test between two neighboring datasets $D \sim D'$. Following \cite{dong2019gaussiandifferentialprivacy}, privacy can be characterized via the hypothesis-testing view through a trade-off function. Formally, let $P$ and $Q$ denote the output distributions of a randomized mechanism $\mathcal{M}$ on $D$ and $D'$, respectively. For any rejection rule testing $H_0:P$ in favor of $H_1:Q$, the trade-off function $T_{(P,Q)}:[0,1]\to[0,1]$ is defined as:
\begin{equation}
T_{(P,Q)}(\alpha) \;=\; \inf \{\beta_\phi : \alpha_\phi\leq \alpha\}, \qquad \alpha\in[0,1],  
\end{equation}
where $\alpha_\phi = \mathbb{E}(\phi)$ and $\beta_\phi=1-\mathbb{E}(\phi)$ are type I and type II errors, respectively. Evidently, larger values of the trade-off function indicate a harder hypothesis testing problem (hence more private).
\begin{definition}[GDP]
A randomized algorithm $\mathcal{M}$ satisfies $\mu$-Gaussian Differential Privacy if for every pair of neighboring datasets $D\sim D'$ with output distributions $(P,Q)$,
\begin{equation}
T_{(P,Q)}(\alpha)\;\ge\; G_{\mu}(\alpha):= \Phi\!\big(\Phi^{-1}(1-\alpha)-\mu\big), \qquad \forall\,\alpha\in[0,1].
\end{equation}
where \(G_{\mu}(\alpha)\) is the benchmark trade-off curve for testing $\mathcal{N}(0,1)$ against $\mathcal{N}(\mu,1)$, and $\mu$-GDP asserts that distinguishing $\mathcal{M}(D)$ from $\mathcal{M}(D')$ is no easier than the Gaussian benchmark.
\end{definition}

DP has its limitations: strong protection often entails utility loss, and its guarantees are typically uniform across all users and records, neglecting that not all secrets are equally sensitive. As a result, DP can be overly conservative for secret protection.

\paragraph{DP Synthetic Text Generation.}
The goal of generating DP synthetic texts is to mimic private data while protecting private information from leakage. An intuitive way is to train a language model with DP-SGD~\cite{abadi2016deep} as a generator to guarantee DP for private data. While DP-Generator~\cite{yue2023synthetic,yu2024privacypreserving,tan2025synthesizing} is effective, it is computationally intensive and requires a large amount of high-quality private data to achieve strong performance. Furthermore, it cannot benefit from state-of-the-art LLMs, as these models (e.g., GPT, Claude, Gemini, etc.) are closed-source, which limits its potential. To handle the above issues, Private Evolution (PE)~\cite{xie2024differentiallyprivatesyntheticdata} is proposed for DP text synthesis, which only requires API access to foundation models and iteratively updates randomly initialized samples. Then,~\cite{zou2025contrastive} and~\cite{hou2025private} further extend PE to data-deficient and data-isolated scenarios. However, standard DP provides uniform protection across all data, which can lead to excessive utility loss and computational overhead, especially when only a small portion of data is truly sensitive.
\section{Method}

To reduce the amount of noise added and thereby improve utility, we adopt secret protection in place of differential privacy. The formal definition of secret protection is provided in Section~\ref{sec:secret-protect}. Building on this definition, we introduce Secret-Protected Evolution (SecPE) in Section~\ref{sec:secpe}. In particular, Section~\ref{sec:secret-noise} derives the noise scale required to achieve secret protection. Given this noise scale, Section~\ref{sec:secret-cluster} presents a private clustering method that satisfies the definition of secret protection, while significantly reducing computational complexity compared to Private Evolution. Finally, Section~\ref{sec:secpe-pipeline} summarizes the SecPE pipeline and establishes the privacy guarantee of the SecPE algorithm.

\subsection{Secret protection}
\label{sec:secret-protect}
The notion of secrets can be subjective, as it is determined by individuals' subjective will. On one hand, secrets may be proprietary business information that an organization seeks to protect (e.g., data containing trading details from an investment firm). On the other hand, secrets may comprise user information that a company wishes to leverage for model training. In general, secrets refer to sensitive content that warrants protection and the focus is not on membership privacy, but on safeguarding the secrets themselves against reconstruction.
\cite{ganesh2025hushprotectingsecretsmodel} introduces a framework that provides privacy guarantees calibrated to the sensitivity of each secret, in contrast to DP, which enforces a uniform and often overly conservative level of protection.
\begin{definition}[Secret Protection]
Let \( D = \{x_1, \dots, x_n\} \) be a training dataset, where each sample may contain secrets from \( S = \{s_1, \dots, s_m\} \). For a secret \( s_j \in S \), let \( \pi_j \) denote a prior distribution over datasets \(\{D_j^1, \dots, D_j^K\}\) such that $\Pr(D_j^k) \leq p_j$, where \( D \) and \( D_j^k \) differ exclusively in the presence of \( s_j \). A randomized mechanism \( \mathcal{A} \) is said to satisfy \( (\vp,\vr) \)-secret protection if, for any reconstruction attack \( B \), the following holds:
\begin{equation}
    \Pr_{D_{j} \sim \pi_j,\, \mathcal{A}} \left[ B\big(\mathcal{A}(D_{j})\big) = s_j \right] \leq r_j, \quad \forall j.
\end{equation}
\end{definition}
Here, $\vp$ and $\vr$ are vectors, and \( \pi_j \) encodes the adversary's prior knowledge about the secret \( s_j \). The guarantee bounds, for each secret, the posterior reconstruction probability given a specified prior success probability. This notion closely parallels the \emph{reconstruction robustness} of \cite{balle2022reconstructingtrainingdatainformed}, where sharp bounds are obtained via the blow-up function. Accordingly, we interpret \((\vp, \vr)\)-secret protection through the lens of $\mu$-GDP, since the trade-off function is tightly connected to blow-up function. We align the neighboring relation so that $D \simeq D_j$ differ only by a single element:  specifically, $s_j \in x \in D$ and $s_j \notin x^\prime \in D_j$.

\begin{lemma}
\label{lem:mugdp_to_secdp}
Any $\mu$-GDP mechanism $\mathcal{A}$ provides \((\vp, \vr)\)-secret protection, where
\begin{align}
\label{eq:gdp-secret}
r_j \;=\; 1 - \Phi\!\left(\Phi^{-1}(1-p_j) - \mu\right).
\end{align}
\begin{proof}
Let $P$ and $Q_j$ denote the distributions of $\mathcal{A}(D)$ and $\mathcal{A}(D_j)$, respectively.  By $\mu$-GDP, we have
\begin{equation}
    \begin{aligned}
    & T_{(P, Q_j)}(p_j) \;\geq\; G_{\mu}(p_j) \;=\; \Phi\!\left(\Phi^{-1}(1-p_j) - \mu\right)\\
    \Longleftrightarrow & 1 - B_{(P, Q_j)}(p_j) \geq \Phi(\Phi^{-1}(1-p_j) - \mu) \\
    \Longleftrightarrow & B_{(P, Q_j)}(p_j) \leq 1 - \Phi(\Phi^{-1}(1-p_j) - \mu) \triangleq r_j
\end{aligned}
\end{equation}
where $T_{(P,Q)}(\alpha) = \inf_{E: Q(E)\leq \alpha} P(E)$ is the trade-off function and 
$B_{(P,Q)}(\alpha) = \sup_{E: Q(E)\leq \alpha} P(E)$ is the blow-up function. 
By Theorem~2 of \cite{hayes2023boundingtrainingdatareconstruction}, this directly implies $(\vp,\vr)$-secret protection.
\end{proof}
\end{lemma}
\vspace{-2pt}
We use GDP to interpret secret protection because the posterior obtained from the trade-off curve yields a tight single-point bound at the specified prior. Note that the converse does not hold in general: $(\vp,\vr)$-secret protection constrains the adversary’s success at a single prior $p_j$, 
whereas $\mu$-GDP requires the entire trade-off curve $T_{(P,Q)}$ to lie above $G_\mu$. This highlights that $(\vp,\vr)$-secret protection constitutes a relaxation of the classical DP definition.

\subsection{Secret-Protected Evolution}
\label{sec:secpe}
In this section, we propose \textbf{Sec}ret-\textbf{P}rotected \textbf{E}volution (SecPE), a framework that incorporates \emph{secret protection} into the evolution paradigm introduced in prior work. 
Whereas traditional PE~\cite{lin2025differentiallyprivatesyntheticdata,
xie2024differentiallyprivatesyntheticdata,
yu2024privacypreserving} operate under canonical DP guarantees, SecPE shifts the focus to \emph{secret protection}, aligning with the discussion in the previous subsection.  

Traditional PE draws random samples from a foundation model and iteratively refines candidates through DP voting on similarity to the private data. A key drawback of this approach is the high computational cost from redundant similarity evaluations. Moreover, the voting distribution is typically unbalanced, with a small subset of synthetic samples accumulating the majority of votes while the rest are selected almost uniformly (see Figure~\ref{fig:vote}). 
This imbalance reveals inefficiencies in the selection procedure and motivates a more structured clustering-based design.  

To address these issues, Algorithm~\ref{alg:secpe} replaces direct voting with cluster representatives. 
At a high level, the SecPE pipeline consists of two stages:  
(1) \emph{Secret Clustering}, where public data are clustered and updated with noisy private contributions to form representative anchors; and  
(2) \emph{Protected Evolution}, where noisy representatives replace individual private samples in the voting process.
This design preserves the practicality of PE while explicitly tailoring protection to secrets.

Specifically, for $M$ private examples and a target of $N_{\mathrm{syn}}$ synthetic samples, the naive PE voting scheme requires $O(MN_{\mathrm{syn}})$ similarity computations. 
In contrast, SecPE leverages Secret Clustering with $K$ anchors, reducing the complexity to $O(KN_{\mathrm{syn}})$, where typically $K \ll M$. 
This yields substantial runtime savings in practice, as further validated in Table~\ref{tab:time}.

\subsubsection{Noise for Secret Protection} 
\label{sec:secret-noise}
Following \cite{ganesh2025hushprotectingsecretsmodel}, we assign a weight $w_i$ to each private example via linear program:
\begin{equation}
\label{eq:secret-noise}
\max_{x_i \in D_{\mathrm{pri}}} \; w_i 
\quad \text{subject to} \
\sum_{x_i \in D_{\mathrm{pri},j}} w_i \;\le\; 
\Phi^{-1}(1-p_j) - \Phi^{-1}(1-r_j) \;\triangleq\; \eta_j,
\quad w_i \in [0,1]\ \ \forall i.
\end{equation}
where \(D_{\mathrm{pri},j} := \{\,x_i \in D_{\mathrm{pri}} \mid s_j \in x_i\,\}\) is a subset of $D_{\mathrm{pri}}$ such that each data in $D_{\mathrm{pri},j}$ contains secret $s_j$. The objective encourages including as many examples as possible; We then construct sampling probabilities $\rho_i = \tfrac{1}{\max_i w_i}\tfrac{w_i}{\sum_{i^\prime}w_{i^\prime}}$ to form a training subset.
Here, \(\eta_j = \mu\) in Equation~\ref{eq:gdp-secret}  acts as a natural capacity constraint, permitting more samples to be selected when \(s_j\) is less sensitive (i.e., larger $\eta_j$). However, secret protection only requires an \emph{upper bound} on the blow-up function. In practice, \(\eta_j\) may be chosen heuristically. A detailed procedure is outlined in Algorithm~\ref{algo: secret noise}.

\begin{algorithm} [H]
\caption{Procedure \textsc{SecretNoise}}
\label{algo: secret noise}
\begin{algorithmic}[1]
\STATE {\bfseries Input:} Dataset $D_{\mathrm{pri}}$, secrets $S_{\mathrm{sec}}$, secret budget $(\vp, \vr)$.
\STATE {\bfseries Output:} noise parameter $\sigma$, sampling probabilities $\bm{\rho}$
    \STATE $\{w_i\} \leftarrow$ solution to linear program (\ref{eq:secret-noise}) using the chosen $\eta_j$ values.
    \STATE $V \gets 1/\max_i w_i$, $\rho_i \gets  V\cdot \tfrac{w_i}{\sum_{i^\prime} w_{i^\prime}}$
     \STATE For each $s_j \in S_{\mathrm{sec}}$,  $P_j = \mathcal{N}(\sum_{x_i \in D_{\mathrm{sec},j}} \mathrm{Bern}(\rho_i), \sigma^2)^{\otimes T},\ Q_j = \mathcal{N}(0, \sigma^2)^{\otimes T}$
     \STATE $\sigma_j \gets \min\{\sigma: B_{(P_j,Q_j)}(p_j)\leq r_j\}, \quad \sigma \gets \max_j \sigma_j$
\end{algorithmic}
\end{algorithm}
\vspace{-6pt}

\subsubsection{Secret Clustering} 
\label{sec:secret-cluster}
In the PE procedure, a small number of synthetic samples receive the majority of votes, as illustrated in Figure~\ref{fig:vote}. This phenomenon indicates that selection occurs in group-like clusters rather than being uniformly spread, which motivates replacing pointwise voting with representative voting via clustering.
Predefined secrets allow us to first detect and cluster using only public data, and then apply a controlled shift informed by secret-containing data. In this way, the representative centers summarize the global structure of the dataset without directly exposing sensitive information. Synthetic data can then be selected based on proximity to these representative centers, eliminating the need to repeatedly process the entire dataset, a procedure that is especially costly for large-scale datasets.

A noteworthy hyperparameter is the number of clusters $K$. 
In practice, we recommend scaling $K$ with both the size of the original data and the target number of synthetic samples: 
(1) large enough to support diverse voting, and 
(2) small enough to limit noise amplification. 
Empirically, performance is largely insensitive to the exact choice of $K$ (see Section~\ref{sec:random words}).

\begin{algorithm} [H]
\caption{Procedure \textsc{SecretClustering}}
\label{algo:secret-clustering}
\begin{algorithmic}[1]
\STATE {\bfseries Input:} Dataset $D_{\mathrm{pri}}\cup D_{\mathrm{pub}}$, secrets $S_{\mathrm{sec}}$, text embedding model $\Psi$, clipping radius $R$.
\STATE {\bfseries Input:} Public clusters $\{(e_k,n_k)\}_{k=1}^K = \textsc{Kmeans}(D_{\mathrm{pub}}, K)$.
\STATE {\bfseries Input:} $(\sigma,\ \bm{\rho}) = \textsc{SecretNoise}(D_{\mathrm{pri}},S_{\mathrm{sec}}, \vp, \vr)$.
\STATE {\bfseries Output:} Noisy cluster centers and cluster sizes $\{(\tilde e_k,\tilde n_k)\}_{k=1}^K$.
     \STATE $E_\mathrm{pri}\gets\text{Clip}_R(\Psi (D_{\mathrm{pri}}))$;\ Initialize $e_k = n_k \cdot e_k $, \ $m_k=0$
     \FOR{$e_{\mathrm{pri},i} \in E_{\mathrm{pri}}$}
     \STATE Sample $z \sim \mathrm{Bernoulli}(\rho_i)$.
     \IF{$z=1$} 
        \STATE Assign $e_{\mathrm{pri},i}$ to its nearest public center: $k \gets \arg\min_{j \in [K]} d(e_{\mathrm{pri},i},e_j)$.
        \STATE Update cluster statistics: $e_k \gets e_k + e_{\mathrm{pri},i}$, \; $m_k \gets m_k + 1$.
    \ENDIF
    \ENDFOR
\STATE $\tilde{n}_k \gets n_k + m_k+\mathcal{N}(0, \sigma^2)$,\quad $\tilde{e}_k \gets \frac{e_k}{n_k+m_k} + \frac{2R}{n_k }\cdot \mathcal{N}(0, \sigma^2 \mathbb{I}_d)$.
\end{algorithmic}
\end{algorithm}
\vspace{-6pt}
\begin{theorem}[Secret Clustering]
\label{thm:secret-cluster}
Let $\{C_k\}_{k=1}^{K}\triangleq\{(\ve_k,n_k)\}_{k=1}^{K}$ denote the set of public cluster centers with corresponding cluster sizes. Every private vector is clipped as $\hat{\ve}_{\mathrm{pri},i} = \text{Clip}_R(\ve_{\mathrm{pri},i}) = \ve_{\mathrm{pri},i} \cdot \min\{1,R/\Vert\ve_{\mathrm{pri},i} \Vert\}$, and then assigned to its nearest anchor; Let $m_k$ denote the number of private points assigned to anchor $\ve_k$. For every cluster $k$, we release the perturbed statistics:
\begin{equation}
\begin{aligned}
    \tilde{\ve}_k &:= \frac{\,n_k \cdot \ve_k + \sum_{i\in C_k}\hat \ve_{\mathrm{pri},i}}{n_k+m_k} + \xi_k,
\quad
\xi_k\sim\tfrac{2R}{n_k}\cdot\mathcal{N}(0,\sigma^{2}I_d),\\
\tilde n_k  &= n_k + m_k + \eta_k,
\qquad
\eta_k\sim\mathcal{N}(0,\sigma^2).
\end{aligned}
\end{equation}
where $\sigma$ is chosen by Algorithm~\ref{algo: secret noise} with $T=1$. Then Algorithm~\ref{algo:secret-clustering} satisfies $(\vp,\vr)$-secret protection.
\end{theorem}

\subsubsection{SecPE Pipeline}
\label{sec:secpe-pipeline}
Following \cite{xie2024differentiallyprivatesyntheticdata}, Algorithm~\ref{alg:secpe} instantiates SecPE with two interacting components each round:
(i) Secret Clustering via Algorithm~\ref{algo:secret-clustering} to build noisy representatives and voting weights; and
(ii) Protected Evolution that alternates selection with LLM-driven variation.
Specifically, the procedure begins with an initialization step (line~4 in Algorithm~\ref{alg:secpe}) that prompts a foundation model to generate random samples. 
At each iteration, the top $N_{\mathrm{syn}}$ candidates from the previous round are selected based on their similarity to the secret-protected clustering centers. 
These survivors, together with their LLM-generated variations, form the candidate pool for the next round. 
A detailed convergence analysis is deferred to Appendix~\ref{sec:convergence analysis}.

\begin{algorithm} [H]
\caption{SecPE Pipeline}
\begin{algorithmic}[1]
\label{alg:secpe}
\STATE {\bfseries Input:} Dataset $D_{\mathrm{pri}}\cup D_{\mathrm{pub}}$, secrets $S_{\mathrm{sec}}$, text embedding model $\Psi$
\STATE {\bfseries Input:} Number of synthetic samples $N_\mathrm{syn}$, variation number $L$.
\STATE {\bfseries Output:} Synthetic text dataset $S_\mathrm{syn}^{T}$
    \STATE Initialize $S_0 \gets  \mathrm{RANDOM}(N_\mathrm{syn}*L)$
    \FOR{$t \in \{0, 1, \cdots, T-1\}$}   
    \STATE $E_t \gets \Psi (S_t)$,\ $E_t \gets E_t\cdot\min\!\bigl(1,R/\|E_t\|\bigr)$,\ $\mathrm{Histogram}_t\gets [0,\dots,0]$
    \STATE $\{(\tilde{\ve}_k, \tilde{n}_k)\}_{k=1}^{K} \gets \textsc{SecretClustering}(D_{\mathrm{pub}}, D_{\mathrm{pri}})$ 
    \FOR{ $k \in \{1,\dots, K\}$ }
        \STATE $i \gets \arg\min_{j: e_j \in E_t} d(\tilde{\ve}_k,\ \ve_j)$
        \STATE $\mathrm{Histogram}[i] \gets \mathrm{Histogram}[i] + \tilde{n}_k$
    \ENDFOR
    \STATE $S_\mathrm{syn}^{t+1} \gets$ Top $N_\mathrm{syn}$ samples according to $\mathrm{Histogram}_t$.
    \STATE $S_{t+1} \gets [\mathrm{VARIATION}(S_\mathrm{syn}^{t+1}, L),\ S_\mathrm{syn}^{t+1}]$ 
    \ENDFOR
\end{algorithmic}
\end{algorithm}

\begin{theorem}[Privacy Guarantee for Algorithm~\ref{alg:secpe}]
\label{thm:sec-accounting}
Let Algorithm~\ref{alg:secpe} run for $T$ iterations with noise multiplier $\sigma$ as specified in Algorithm~\ref{algo: secret noise}.  
Then it satisfies $(\vp, \vr)$-secret protection.
\end{theorem}
Theorem~\ref{thm:sec-accounting} provides a theoretical guarantee for Algorithm~\ref{alg:secpe}. The proof follows directly from Theorem~2.4 of \cite{doroshenko2022connectdotstighterdiscrete}, 
since the pair of distributions 
\(\mathcal{N}(\sum_{x_i \in D_{\mathrm{pri},j}} \mathrm{Bern}(\rho_i), \sigma^2)\) and 
\(\mathcal{N}(0, \sigma^2)\) form a dominating pair in each round.

\begin{remark}
When cosine similarity is used as the distance metric, it suffices to apply $K$-means to $\ell_2$-normalized embeddings, 
which effectively transforms the clustering into cosine-based grouping. 
In this case, we set the sensitivity bound $R = 1$ when calibrating the noise scale.
\end{remark}

\section{Experiments}
\label{sec:experiment}

\subsection{Setup}
In this section, we empirically evaluate \textbf{SecPE} on text synthesis and privacy protection. 
We first consider a random word task to illustrate SecPE’s ability to generate high-fidelity synthetic text under secret constraints. Then Personally Identifiable Information (PII) task that assesses protection of truly sensitive content, where PII is detected with \href{https://github.com/Sripaad/ai4privacy}{AI4Privacy}. In both tasks, SecPE delivers better utility at lower runtime: secret protection injects less noise than $\mu$-GDP at the same reconstruction budget, and representative voting with clustering cuts computation while preserving fidelity.

\paragraph{Datasets.}
We evaluate on three widely applied open-source datasets: (1) OpenReview \cite{xie2024differentiallyprivatesyntheticdata}: ICLR 2023 paper reviews labeled by research area and recommendation rating; (2) PubMed \cite{yue2023synthetic}: medical paper abstracts; and (3) Yelp \cite{yelpopendataset}: user reviews of businesses labeled by business category and rating.

\paragraph{Baselines.}
Given that \emph{secret protection} is a new concept, we construct a baseline by instantiating the most popular Aug-PE under $\mu$-GDP, with $\mu$ set via Equation \ref{eq:gdp-secret}, and add Gaussian noise calibrated to the voting sensitivity following \cite{xie2024differentiallyprivatesyntheticdata}.
We use SecPE$_{K}$ to denote SecPE with $K$ clusters. Our comparison covers three aspects: (i) \emph{Computational efficiency}: GPU hours for response generation and computation time for counting histogram are reported. (ii) \emph{Downstream performance}: we fine-tune RoBERTa-base \cite{liu2019robertarobustlyoptimizedbert} on synthetic data to classify Yelp ratings/categories and OpenReview recommendations/areas. For PubMed, bert-base-uncased (BERT) \cite{turc2019wellreadstudentslearnbetter} is fine-tuned to report next-word prediction accuracy;  (iii) \emph{Real–synthetic similarity}: we compute FID on text embeddings and provide a comparison of text-length distributions;

\begin{table}[t]
\centering
\caption{Hyperparameter settings for SecPE and Aug-PE across datasets.}
\setlength{\tabcolsep}{12pt}
\scalebox{0.9}{
\begin{tabular}{@{}ccccccc@{}}
\toprule[1pt]
Dataset & $\text{N}_{\mathrm{syn}}$ & Cluster $K$ & $L$ & Iterations & Temperature & Max tokens \\ \midrule
OpenReview & 2000 & 15, 20, 15 & 6 & 10 & 1.2 & 448 \\
PubMed     & 2000 & 2000, 3000, 4000 & 6 & 5  & 1.2 & 448 \\
Yelp       & 5000 & 400, 600, 800     & 6 & 5  & 1.2 & 64  \\
\bottomrule[1pt]
\end{tabular}
}
\label{tab:hyperparams}
\end{table}

\begin{table}[t]
\centering
\caption{LLM generation time and histogram computation time (seconds) for one epoch.}
\setlength{\tabcolsep}{15pt}
\scalebox{0.92}{
\begin{tabular}{@{}ccccccc@{}}
\toprule[1pt]
\multirow{2}{*}{Time (sec)} & \multicolumn{2}{c}{OpenReview} & \multicolumn{2}{c}{PubMed} & \multicolumn{2}{c}{Yelp} \\ \cmidrule(l){2-7} 
 & LLM & Histogram & LLM & Histogram & LLM & Histogram \\ \midrule
Aug-PE & 1698.7 & 126.9 & 828.5 & 32.2  & 347.1 & 30126.4 \\
SecPE & 1693.1 & \textbf{1.5} & 830.8 & \textbf{0.5} & 347.6 & \textbf{2.3} \\
\bottomrule[1pt]
\end{tabular}
}
\label{tab:time}
\end{table}

\begin{table}[!t]
\centering
\caption{Performance comparison of downstream tasks within random words on PubMed.}
\scalebox{0.75}{
\begin{tabular}{@{}cccccccccc@{}}
\begin{tabular}{@{}cccccccccc@{}}
\toprule[1pt]
\multirow{2}{*}{LLM} & \multirow{2}{*}{Method} & \multicolumn{2}{c}{$\vr/ \vp=2$} & \multicolumn{2}{c}{$\vr/ \vp=10$} & \multicolumn{2}{c}{$\vr/ \vp=50$} & \multicolumn{2}{c}{r$/p=\infty$} \\ \cmidrule(l){3-10} 
 &  & BERT\_m & BERT\_s & BERT\_m & BERT\_s & BERT\_m & BERT\_s & BERT\_m & BERT\_s \\ \midrule
\multirow{4}{*}{GPT2} & AugPE & 22.15 & 24.93 & 23.13 & 26.14 & 24.39 & 26.96 & \textbf{27.28} & \textbf{29.70} \\
 \cmidrule(l){2-10} 
 & SecPE$_{2000}$ & 26.74 & 29.18 & 27.09 & 29.42 & 26.82 & 29.38 & 26.82 & 29.19 \\
 & SecPE$_{3000}$ & \textbf{27.15} & 29.54 & \textbf{27.32} & \textbf{29.75} & 26.69 & 29.12 & 27.09 & 29.52 \\
 & SecPE$_{4000}$ & 27.02 & \textbf{29.57} & 26.86 & 29.34 & \textbf{27.23} & \textbf{29.43} & 27.01 & 29.41 \\ \midrule
\multirow{4}{*}{Qwen-2.5-1.5B} & AugPE & 20.37 & 22.65 & 21.01 & 23.09 & 21.18 & 23.52 & \textbf{23.68} & \textbf{25.87} \\
 \cmidrule(l){2-10} 
 & SecPE$_{2000}$ & \textbf{23.17} & \textbf{25.37} & 22.26 & 24.40 & \textbf{22.93} & \textbf{24.96} & 22.50 & 24.63 \\
 & SecPE$_{3000}$ & 22.31 & 24.48 & 22.24 & 24.55 & 22.65 & 24.92 & 22.84 & 24.91 \\
 & SecPE$_{4000}$ & 22.17 & 24.41 & \textbf{22.63} & \textbf{24.99} & 22.84 & 24.91 & 22.29 & 24.40 \\ \bottomrule[1pt]
\end{tabular}
\end{tabular}
}
\label{tab:pubmed_random}
\end{table}

\paragraph{Implementation details.}
As text generators, we use GPT-2 \cite{radford2019language}, Qwen-2.5-1.5B \cite{qwen2025qwen25technicalreport} for main experiments. Llama-3.1-8B \cite{dubey2024llama3}, Qwen-2.5-7B \cite{qwen2025qwen25technicalreport}, Mistral-7B-Instruct-v0.3 \cite{jiang2023mistral7b} and GPT-4o-Mini \cite{gpt4o} are applied for ablation study. 

We use Sentence-Transformers \cite{reimers2019sentencebertsentenceembeddingsusing} as the embedding model $\Psi$. For the privacy budget, we fix the prior vector $\boldsymbol{p} = \mathbf{1}\!\cdot\!10^{-4}$ and set the budget via the ratio 
$\boldsymbol{r}/\boldsymbol{p} = c$ with $c \in \{2,10,50,\infty\}$, 
where $c=\infty$ denotes the non-private setting. Although one could carefully tailor heterogeneous, secret-specific budgets to achieve better effectiveness, we adopt a uniform budget to enable a fair comparison with the $\mu$-GDP Aug-PE baseline, where $
\mu \;=\; \Phi^{-1}(1-\boldsymbol{p}) \;-\; \Phi^{-1}(1-\boldsymbol{r})$. For numerical stability, we approximate Mixture of Gaussian with a single Gaussian in both settings. Additional training and hyperparameter details are provided in Table~\ref{tab:hyperparams}. For each dataset, the hyper-parameters are kept fixed across all methods. For generating LLMs' responses, the prompts are the same as~\cite{xie2024differentiallyprivatesyntheticdata}.

\subsection{Performance Comparison on Random Words}
\label{sec:random words}
In this task, we sort all vocabulary items in each dataset by frequency and designate words near the 20\% quantile as secrets; a sample is treated as secret-containing if it includes any designated word.

\paragraph{Runtime comparison.} 
Table~\ref{tab:time} reports runtime on a NVIDIA A100 (80\,GB) GPU. 
A key advantage of SecPE lies in its efficiency: \emph{Secret Clustering} drastically reduces per-iteration histogram construction and selection time, accelerating the overall pipeline. 
Among Experiments, our method reduces this component by at least a factor of $60\times$; on Yelp (1.9M records), the reduction reaches roughly $10{,}000\times$. In terms of LLM sampling, SecPE achieves runtime comparable to Aug-PE, as both methods query the same model for the same number of variations.

\paragraph{Downstream Task.} 
Experimental results comparing SecPE with Aug-PE on downstream tasks are reported in Tables~\ref{tab:pubmed_random}, \ref{tab:openreview_random} and \ref{tab:yelp_random}. For each privacy budget and model, we highlight the highest classification accuracy in \textbf{bold}. On PubMed, as $\vr/ \vp$ decreases from $\infty$ to $2$, the BERT-small next-word prediction accuracy on Aug-PE (GPT-2) synthetic text drops from $29.70\to24.93$,  whereas only a marginal change from $29.19\to29.18$ for $\text{SecPE}_{2000}$.
The results show that, with the same number of training epochs, SecPE consistently achieves higher accuracy under private settings, and this advantage becomes more pronounced as the privacy requirement tightens (i.e., smaller $\vr/ \vp$). 
In the non-private case ($\vr/ \vp=\infty$), performance is slightly lower but broadly comparable, likely because clustering abstracts away fine-grained details and can occasionally induce mis-selections. 
We further observe non-systematic fluctuations in downstream accuracy when varying $K$, attributable to randomness in both the SecPE procedure and the downstream fine-tuning; overall, however, the method remains robust and not sensitive to the choice of $K$.

\begin{table}[t]
\centering
\caption{Performance comparison of downstream tasks within random words on OpenReview.}
\setlength{\tabcolsep}{9.5pt}
\scalebox{0.8}{
\begin{tabular}{@{}cccccccccc@{}}
\toprule[1pt]
 &  & \multicolumn{2}{c}{$\vr/ \vp=2$} & \multicolumn{2}{c}{$\vr/ \vp=10$} & \multicolumn{2}{c}{$\vr/ \vp=50$} & \multicolumn{2}{c}{$\vr/ \vp=\infty$} \\ \cmidrule(l){3-10} 
\multirow{-2}{*}{LLM} & \multirow{-2}{*}{Method} & Area & Rating & Area & Rating & Area & Rating & Area & Rating \\ \midrule
 & Aug-PE & 29.06 & 25.70 & 27.94 & 27.12 & 32.48 & 27.88 & \textbf{41.06} & 28.70 \\
 \cmidrule(l){2-10} 
 & SecPE$_{15}$ & \textbf{30.77} & 30.26 & 31.88 & \textbf{30.70} & 30.34 & 28.27 & 39.02 & 28.38 \\
 & SecPE$_{20}$ & 28.98 & \textbf{31.38} & 32.67 & 28.23 & 30.30 & 29.56 & 38.74 & \textbf{30.49} \\
\multirow{-4}{*}{GPT2} & SecPE$_{25}$ & 30.34 & 29.24 & \textbf{34.81} & 30.66 & 32.48 & \textbf{30.31} & 38.60 & \textbf{30.49} \\ \midrule
 & Aug-PE & 32.70 & 25.55 & 32.23 & 25.80 & 36.49 & 28.52 & 40.20 &  28.09 \\
  \cmidrule(l){2-10} 
 & SecPE$_{15}$ & 38.34 & \textbf{27.73} & \textbf{38.67} & 26.02 & 36.09 & \textbf{30.95} & 36.03 & \textbf{32.03} \\
 & SecPE$_{20}$ & 37.17 & 26.94 & 36.95 & 26.44 & 35.85 & 29.52 & 39.63 & 28.30 \\
\multirow{-4}{*}{Qwen-2.5-1.5B} & SecPE$_{25}$ & \textbf{38.92} & 27.66 & 37.03 & \textbf{27.82} & \textbf{40.24} & 28.81 & \textbf{40.24} & 28.86 \\ \bottomrule[1pt]
\end{tabular}
}
\label{tab:openreview_random}
\end{table}

\begin{table}[t]
\centering
\caption{Performance comparison of downstream tasks within random words on Yelp.}
\scalebox{0.8}{
\begin{tabular}{@{}cccccccccc@{}}
\toprule[1pt]
 &  & \multicolumn{2}{c}{$\vr/ \vp=2$} & \multicolumn{2}{c}{$\vr/ \vp=10$} & \multicolumn{2}{c}{$\vr/ \vp=50$} & \multicolumn{2}{c}{$\vr/ \vp=\infty$} \\ \cmidrule(l){3-10} 
\multirow{-2}{*}{LLM} & \multirow{-2}{*}{Method} & Category & Rating & Category & Rating & Category & Rating & Category & Rating \\ \midrule
 & Aug-PE & 71.53 & 47.02 & 71.62 & 54.72 & 72.60 & 64.02 & 73.54 & \textbf{65.28} \\
  \cmidrule(l){2-10} 
 & SecPE$_{400}$ & 72.06 & 61.44 & 72.90 & 58.92 & 72.18 & 64.30 & 72.50 & 61.28 \\
 & SecPE$_{600}$ & 71.96 & 60.38 & 73.82 & 58.36 & \textbf{73.61} & \textbf{65.08} & 72.96 & 62.22 \\
\multirow{-4}{*}{GPT2} & SecPE$_{800}$ & \textbf{72.74} & \textbf{62.46} & \textbf{74.28} & \textbf{63.70} & 73.12 & 63.82 & \textbf{73.58} & 62.48 \\ \midrule
 & Aug-PE & 72.70 & 55.84 & 72.14 & 53.52 & 71.93 & 55.54 & 72.14 & \textbf{59.02} \\
  \cmidrule(l){2-10} 
 & SecPE$_{400}$ & 73.97 & 55.80 & 72.00 & 57.80 & 73.28 & 57.70 & \textbf{73.84} & 58.78 \\
 & SecPE$_{600}$ & 72.22 & 56.04 & \textbf{73.24} & 57.80 & \textbf{73.66} & 58.53 & 73.00 & 58.61 \\
\multirow{-4}{*}{Qwen-2.5-1.5B} & SecPE$_{800}$ & \textbf{74.12} & \textbf{58.64} & 72.60 & \textbf{58.66} & 72.54 & \textbf{60.93} & 73.14 & 57.06 \\ \bottomrule[1pt]
\end{tabular}
}
\label{tab:yelp_random}
\end{table}

\begin{figure}[t]
    \centering
    \includegraphics[width=\linewidth]{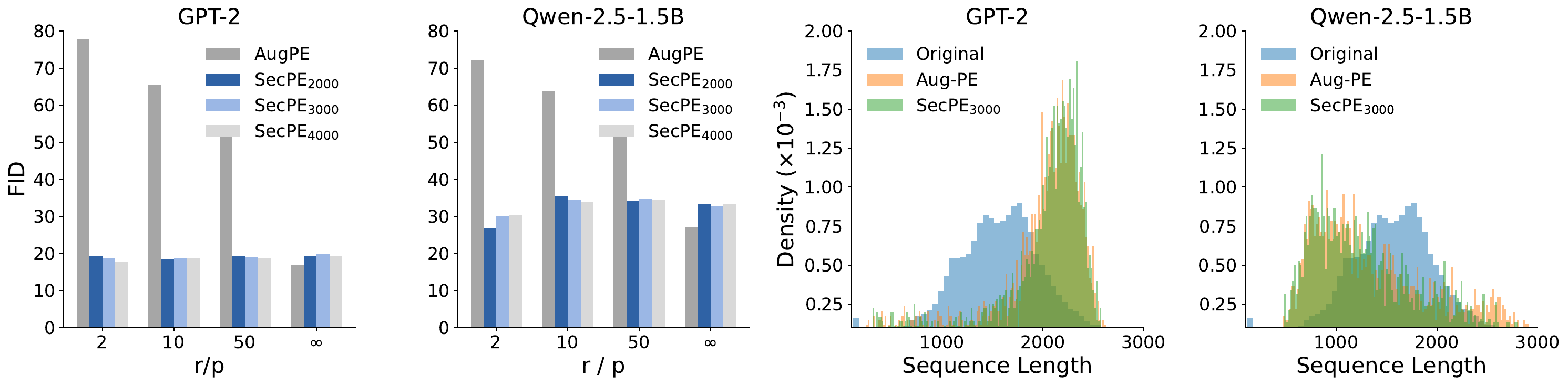}
    \caption{Results on PubMed. (Left) FID relative to the original data for SecPE and Aug-PE under $\vr/ \vp \in \{2,10,50,\infty\}$ using GPT-2 and Qwen-2.5-1.5B. (Right) Synthetic sequence-length distributions for the non-private $\text{SecPE}_{3000}$ and Aug-PE generated by GPT-2 and Qwen-2.5-1.5B, compared with the original data.}
    \label{fig:fid and len pubmed}
\end{figure}

\textbf{Real–synthetic similarity.} The left two panels of Figure~\ref{fig:fid and len pubmed} align with the tabular results:  as $\vr/ \vp$ decreases, SecPE achieves lower FID (i.e., greater similarity to the original data) than Aug-PE, whereas in the non-private setting it yields higher FID. Moreover, FID varies little across all tested $K$, further indicating that the choice of $K$ does not materially affect performance. In the right two panels of Figure~\ref{fig:fid and len pubmed}, we compare the empirical sequence-length distributions of synthetic data with those of the original corpus. As SecPE and Aug-PE rely on the same generative LLM, their length distributions are very similar.

\paragraph{Ablation of LLM.}
We evaluate $\text{SecPE}_{600}$ on Yelp under $\vr/ \vp\in\{10,\infty\}$ using more advanced API-accessible LLMs. As shown in Table~\ref{tab:APIs}, these models achieve accuracy comparable to, or exceeding GPT-2 and Qwen-2.5-1.5B, indicating that our approach benefits from high-quality synthetic text produced by stronger LLMs. 
Within the same family, stronger variants tend to perform better. 
For example, at $\vr/ \vp=10$ (category, rating)-classification accuracy, 
GPT-4o-mini $(74.84,\ 62.96)$ outperforms GPT2 $(73.82,\ 58.36)$, and 
Qwen-2.5-7B $(74.56,\ 63.06)$ outperforms Qwen-2.5-1.5B $(73.12,\ 62.08)$.
However, the 7B Mistral $(72.52,\ 58.10)$ does not outperform a smaller LLM in our setting, suggesting that appropriate model selection, rather than parameter count alone, is crucial.

\begin{table}[t]
\centering
\caption{Stronger LLM generators yield improved downstream accuracy on Yelp.}
\setlength{\tabcolsep}{16pt}
\scalebox{0.88}{
\begin{tabular}{@{}ccccc@{}}
\toprule[1pt]
\multirow{2}{*}{LLM} & \multicolumn{2}{c}{$\vr/ \vp=10$} & \multicolumn{2}{c}{$\vr/ \vp=\infty$} \\ \cmidrule(l){2-5} 
 & Category & Rating & Category & Rating \\ \midrule
GPT2 & 73.82 & 58.36 & 72.96 & 62.22 \\
Qwen-2.5-1.5B & 73.24 & 57.80 & 73.00 & 58.61 \\
\midrule
\midrule
Mistral-7B-Instruct-v0.3 & 72.52 & 58.10 & 73.38 & 61.28 \\
Llama-3.1-8B & 74.14 & 61.92 & 73.82 & 62.99 \\
Qwen-2.5-7B & 74.56 & \textbf{63.06} & 74.24 & \textbf{63.34} \\
GPT-4o-Mini & \textbf{74.84} & 62.96 & \textbf{75.10} & 63.28 \\ \bottomrule[1pt]
\end{tabular}
}
\label{tab:APIs}
\end{table}

\begin{table}[!t]
\centering
\caption{Performance comparison of downstream tasks within PII on Yelp.}
\label{tab:yelp_pii}
\scalebox{0.82}{
\begin{tabular}{@{}cccccccccc@{}}
\toprule[1pt]
\multirow{2}{*}{LLM} & \multirow{2}{*}{Method} & \multicolumn{2}{c}{$\vr/ \vp=2$} & \multicolumn{2}{c}{$\vr/ \vp=10$} & \multicolumn{2}{c}{$\vr/ \vp=50$} & \multicolumn{2}{c}{$\vr/ \vp=\infty$} \\ \cmidrule(l){3-10} 
 &  & Category & Rating & Category & Rating & Category & Rating & Category & Rating \\ \midrule
\multirow{2}{*}{GPT2} & Aug-PE & 73.50 & 62.36 & 73.60 & \textbf{65.89} & {74.10} & \textbf{63.44} & \textbf{73.54} & \textbf{65.28} \\
 & SecPE$_{600}$ & \textbf{73.65} & \textbf{63.45} & \textbf{75.05} & 61.22 & \textbf{74.34} & 62.45 & 72.96 & 62.22 \\ \midrule
\multirow{2}{*}{Qwen-2.5-1.5B} & Aug-PE & 72.97 & 58.30 & 72.70 & 60.04 & 72.83 & 58.93 & 72.14 & \textbf{59.02} \\
 & SecPE$_{600}$ & \textbf{73.00} & \textbf{61.86} & \textbf{72.87} & \textbf{60.60} & \textbf{73.04} & \textbf{59.20} & \textbf{73.00} & 58.61 \\ \bottomrule[1pt]
\end{tabular}
}

\end{table}

\subsection{Performance Comparison on PII}
On the Yelp dataset, we detect 36 PII categories (e.g., age, email, gender) using \href{https://github.com/Sripaad/ai4privacy}{AI4Privacy} and treat each category as a secret, yielding a dense secret-containing corpus. 
In this setting, improvements (see Table~\ref{tab:yelp_pii}) over Aug-PE are modest and less pronounced than in the random-word task. 
It is worth noting, however, that our comparison fixes the number of epochs across methods and therefore does not leverage SecPE’s faster iteration speed. 

\section{Conclusion}
We introduced SecPE, a secret-aware evolution framework for privacy-preserving text synthesis. By calibrating protection at the level of secrets rather than enforcing uniform DP across all records, SecPE provides formal $(\vp,\vr)$-secret guarantees and relaxes Gaussian DP to the operative prior point, thereby achieving tighter utility–privacy trade-offs.  
Empirically, across diverse datasets, SecPE improves fidelity and downstream accuracy over GDP Aug-PE baselines under private settings, while also substantially accelerating the pipeline. Ablation studies further show that stronger LLMs consistently yield higher-quality synthetic text, highlighting the critical role of model selection.
Overall, our results suggest that secret-aware mechanisms offer a more practical and effective approach to privacy-preserving text generation than DP, particularly in settings where sensitive content is sparse in type yet repeated across records and thus highly consequential.

\newpage
\section*{Reproducibility Statement}
For the theoretical analysis of this work, we provide detailed explanations in Section~\ref{sec:secpe}. The assumptions and complete proofs of the theorems are presented in Appendix~\ref{sec:add thms}. 
The datasets, models and hyperparameters used in the experiments are described in Section~\ref{sec:experiment}, and additional experimental settings are reported in Appendix~\ref{sec:supp exps}. All datasets used are publicly available, and their usage are explicitly referenced in the paper.

\section*{Ethics Statement}
This work adheres to the ICLR Code of Ethics, with no involvement of human subjects or animal experimentation. All datasets employed were sourced in compliance with relevant usage guidelines to ensure privacy protection. We ensured the avoidance of biases or discriminatory outcomes throughout the research process, utilized no personally identifiable information, and conducted no experiments posing privacy or security risks. Furthermore, we are committed to upholding transparency and integrity in the research. The proposed method facilitates privacy protection, and we aim to further standardize the use of private data.

\bibliography{ref.bib}
\bibliographystyle{plain}

\newpage
\appendix
\section*{The Use of Large Language Models (LLMs)}
In this work, Large Language Models (LLMs) are used solely as general-purpose assistive tools to
help polish and improve the clarity of the writing. Authors take full responsibility for all content in
the paper, including text that was refined using LLMs, and confirm that no part of the manuscript
generated by LLMs constitutes plagiarism or scientific misconduct.

\section{Limitation}
While clustering accelerates the pipeline, it abstracts away fine-grained details, which can cause a modest loss of utility in the non-private regime. 
Another open challenge is the formal definition of what constitutes a secret and how to quantify its sensitivity. 
Future work will explore heterogeneous, secret-specific budgets and adaptive priors to further improve utility while maintaining protection. 
We also plan to extend SecPE to image domains (with an appropriate formalization of ``secrets'' in that setting) and investigate secret-protected generators.

\section{Supplementary Experiments}
\label{sec:supp exps}
\subsection{Simulation Result}
To explicitly demonstrate how secret protection reduces the required noise relative to GDP, 
we consider a toy setup with $N=8000$ records and $m=400$ secrets. 
Each record contains each secret independently with probability $\mathrm{Bernoulli}(0.01)$. 
For a fair comparison, $\mu$ in GDP is coupled to the $(\vp,\vr)$ pair via
\(
\mu \;=\; \Phi^{-1}(1-p)\;-\;\Phi^{-1}(1-r).
\)

We report the noise ratio defined as $\sigma_{\text{GDP}}/\sigma_{\text{secret}}$ (larger is better). 
In the left panel, we fix $(N,m)=(8000,400)$ and vary the privacy budget $r/p \in [2,400]$. 
In the right panel, we fix $r/p=10$ and $N=8000$, and vary the number of secrets $m \in [100,1000]$. 
Across both settings, the noise required by $(p,r)$-secret protection is consistently smaller than that of $\mu$-GDP.

\begin{figure}[htbp]
    \centering
    \vspace{-10pt}
    \subfigure[Privacy budget]{
        \begin{minipage}[H]{0.4\linewidth}
            \centering
            \includegraphics[width=.92\linewidth]{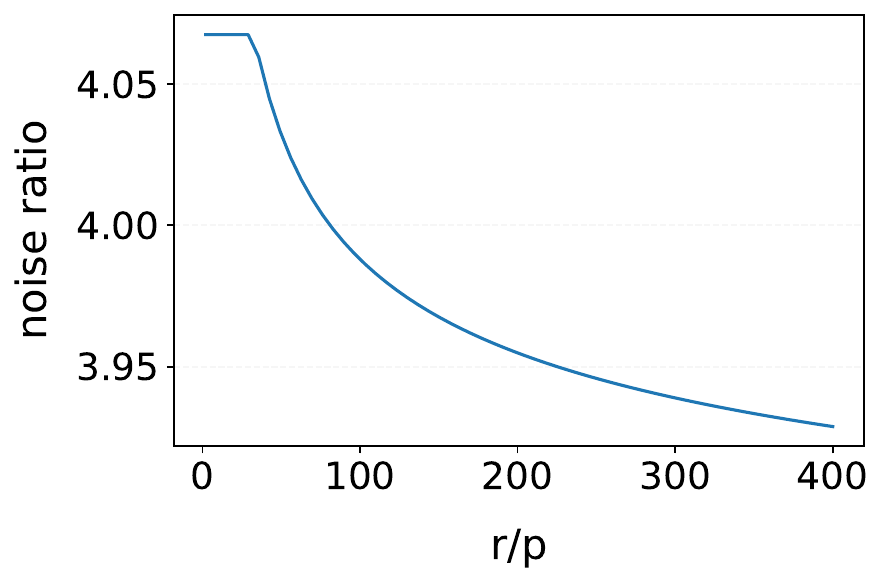}
        \end{minipage}
    }
    \subfigure[Secret fraction]{
        \begin{minipage}[H]{0.4\linewidth}
            \centering
            \includegraphics[width=.92\linewidth]{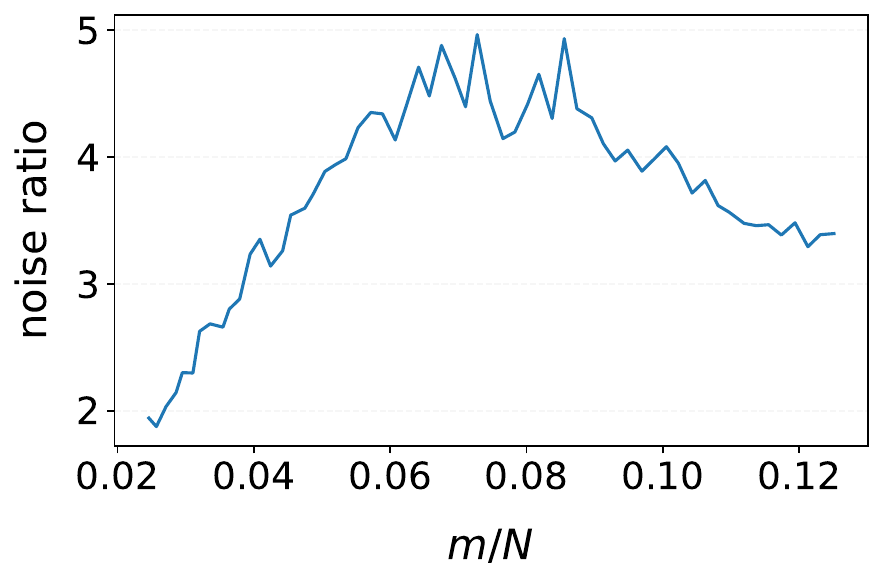}
        \end{minipage}
    }
    \vspace{-8pt}
    \caption{Noise ratio $\sigma_{\text{GDP}}/\sigma_{\text{secret}}$ comparing $(\vp,\vr)$-secret protection with Gaussian DP. (a): $N=8000$, $m=400$, varying $r/p$. (b): $N=8000$, $r/p=10$, varying the number of secrets $m$.}
\end{figure}

\subsection{Implementation Details}
We provide the implementation details of our experiments in \cref{sec:experiment}. For each dataset, the hyperparameters are kept fixed across all methods. For the prompts used to generate responses, we refer readers to \citep{xie2024differentiallyprivatesyntheticdata}.

\subsection{Voting Details}
\begin{figure}[ht]
    \centering
    \vspace{-8pt}
    \subfigure{
    \begin{minipage}[H]{0.33\linewidth}
        \centering
        \includegraphics[width=.92\linewidth]{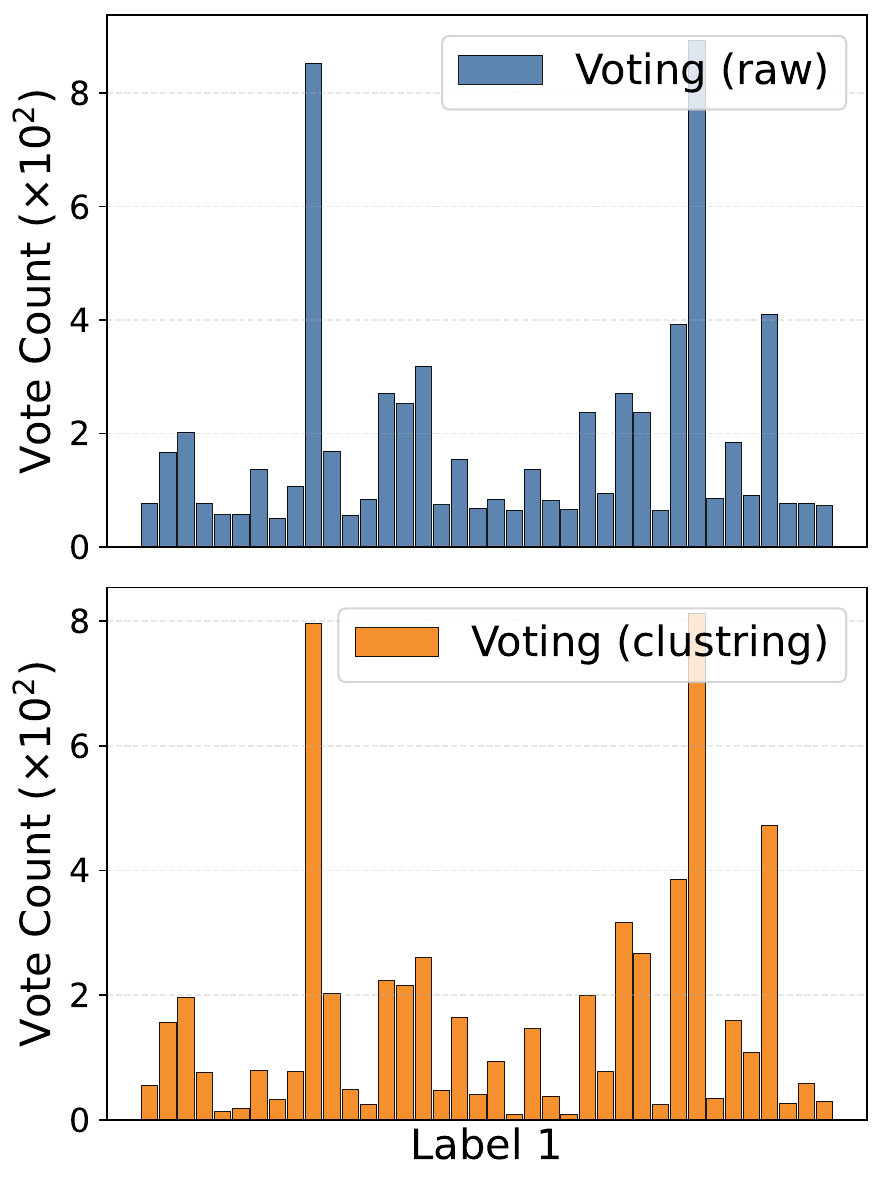}
        \end{minipage}%
    }%
    \subfigure{
    \begin{minipage}[H]{0.33\linewidth}
            \centering
        \includegraphics[width=.92\linewidth]{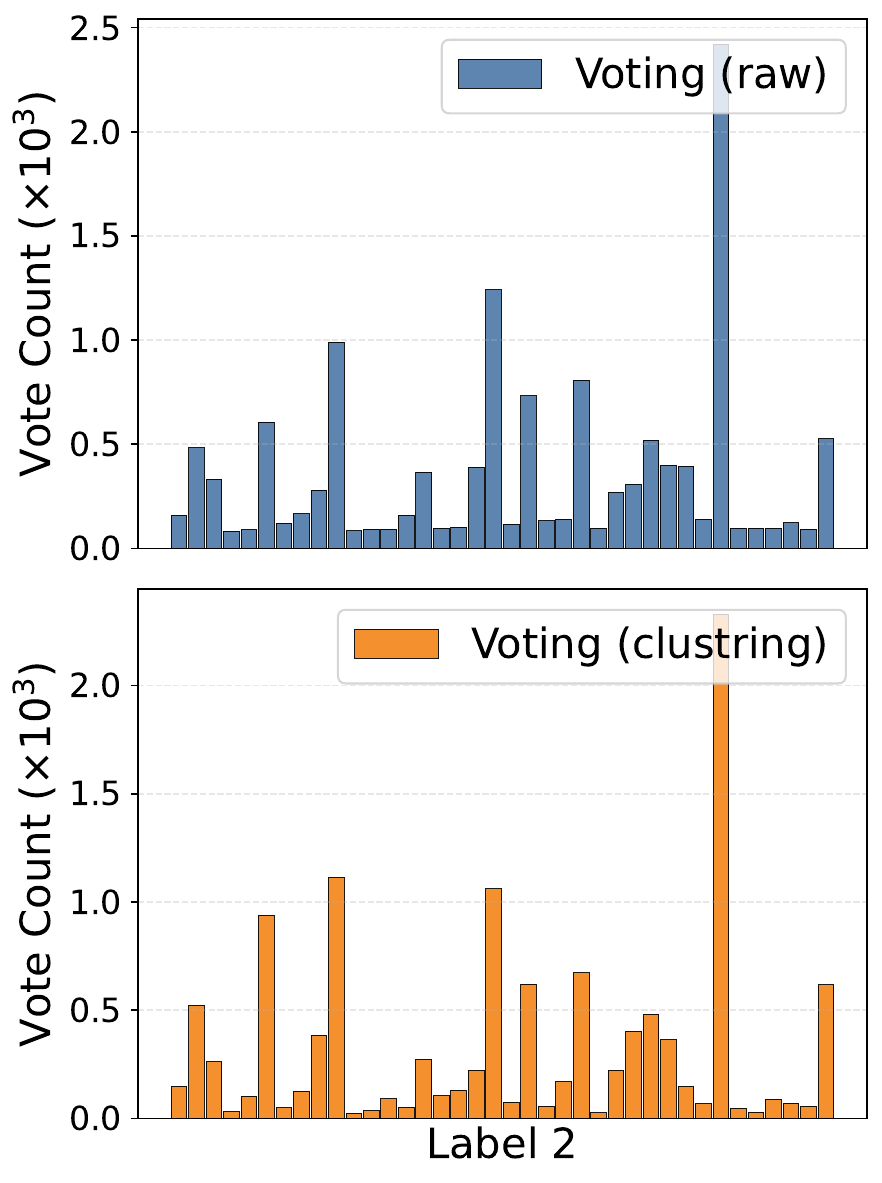}
        \end{minipage}%
    }%
     \subfigure{
    \begin{minipage}[H]{0.33\linewidth}
            \centering
        \includegraphics[width=.92\linewidth]{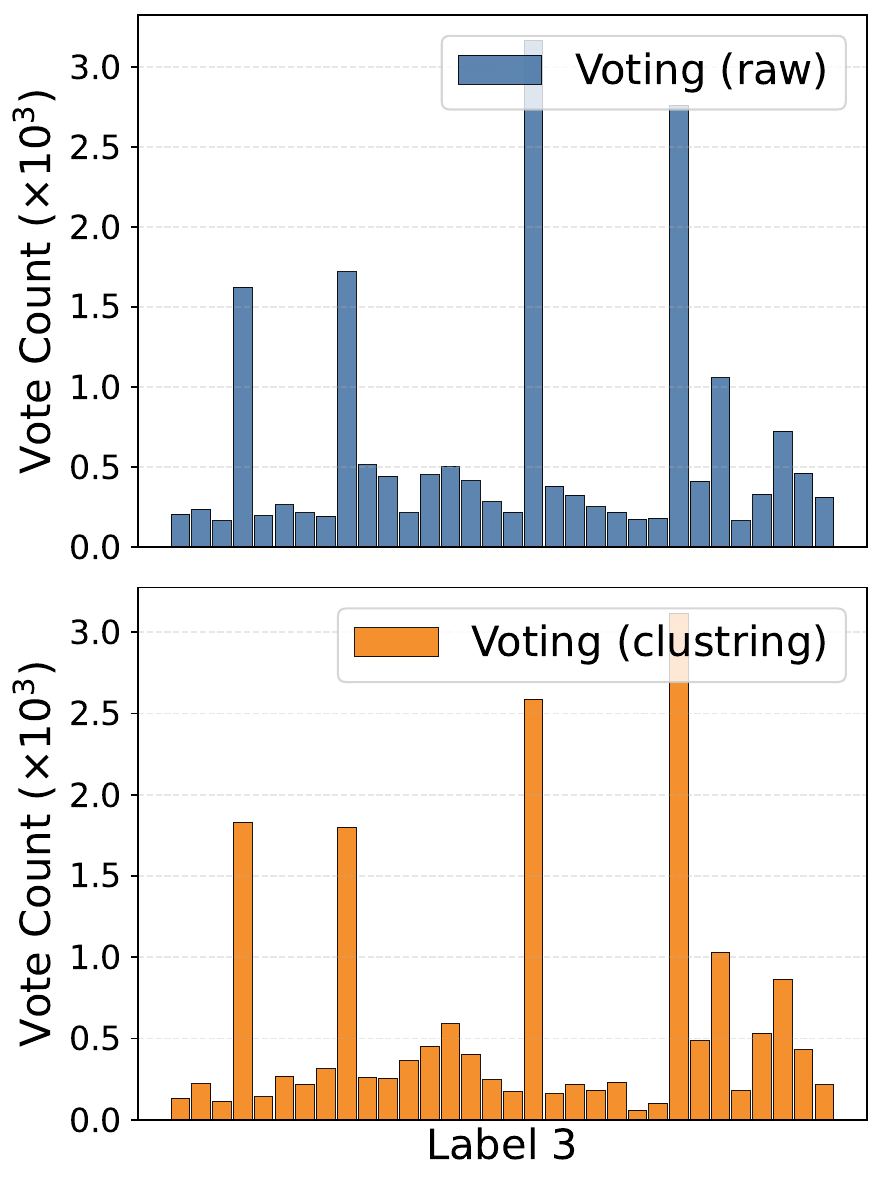}
        \end{minipage}%
    }%
    \vspace{-8pt}
    \caption{Voting distribution per label on Yelp. \emph{Top:} raw votes from Aug-PE. \emph{Bottom:} votes after clustering in SecPE.}
    \label{fig:vote}
\end{figure}

Here we compare raw voting in Aug-PE with post-clustering votes in SecPE. 
Figure \ref{fig:vote} shows the first three labels on Yelp. 
The synthetic samples receiving the highest vote mass are largely the same across both methods, indicating that the clustering preserves the key selections and is thus reasonable.

\subsection{FID and Length}
\begin{figure}[t]
    \centering
    \includegraphics[width=1\linewidth]{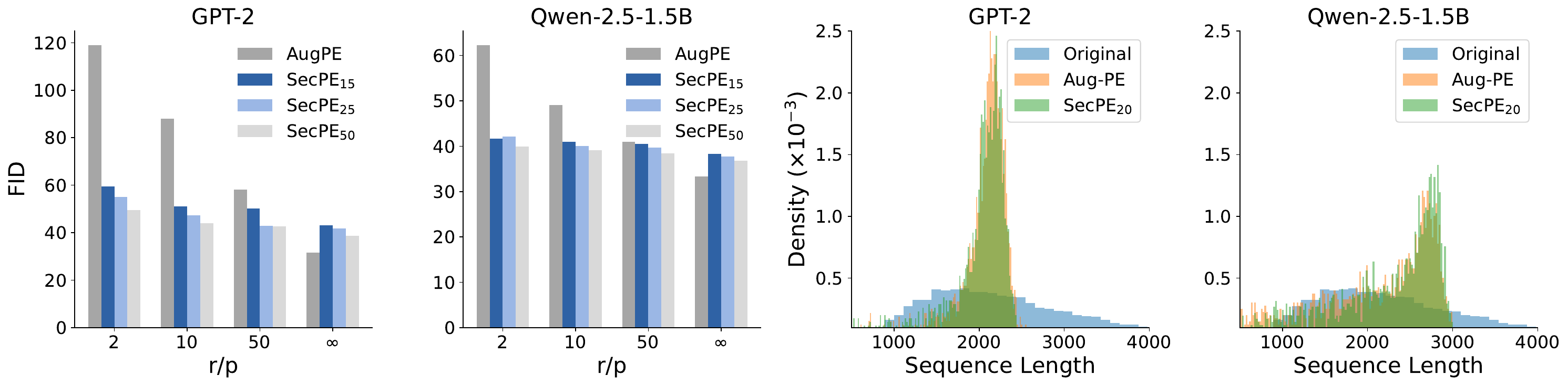}
    \caption{FID and sequence-length distributions on OpenReview.}
    \label{fig:fid and len openreview}
\end{figure}

Figure~\ref{fig:fid and len openreview} presents the FID with respect to the OpenReview dataset for SecPE and Aug-PE under $r/p \in \{2,10,50,\infty\}$, together with the synthetic sequence-length distributions of the non-private $\text{SecPE}_{20}$ compared against Aug-PE, using GPT-2 and Qwen-2.5-1.5B as generators.

\begin{figure}[ht]
    \centering
    \includegraphics[width=\linewidth]{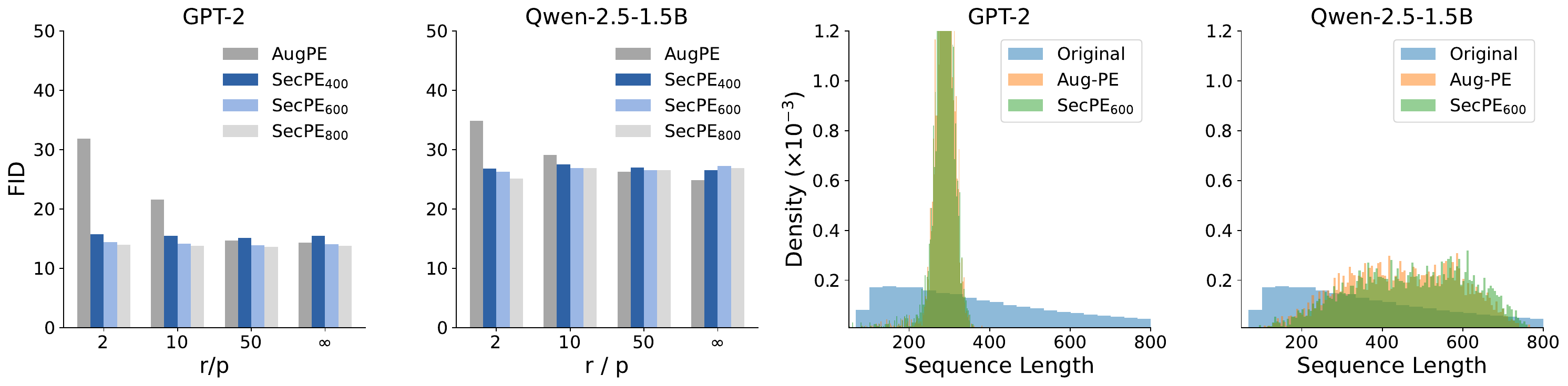}
    \caption{FID and sequence-length distributions on Yelp.}
    \label{fig:fid and len yelp}
\end{figure}

Similarly, Figure~\ref{fig:fid and len yelp} shows the FID with respect to the Yelp dataset for SecPE and Aug-PE under the same $r/p$ settings, along with the sequence-length distributions of the non-private $\text{SecPE}_{600}$ compared against Aug-PE, also using GPT-2 and Qwen-2.5-1.5B as generators.

\subsection{APIs}
\begin{figure}[ht]
    \centering
    \includegraphics[width=\linewidth]{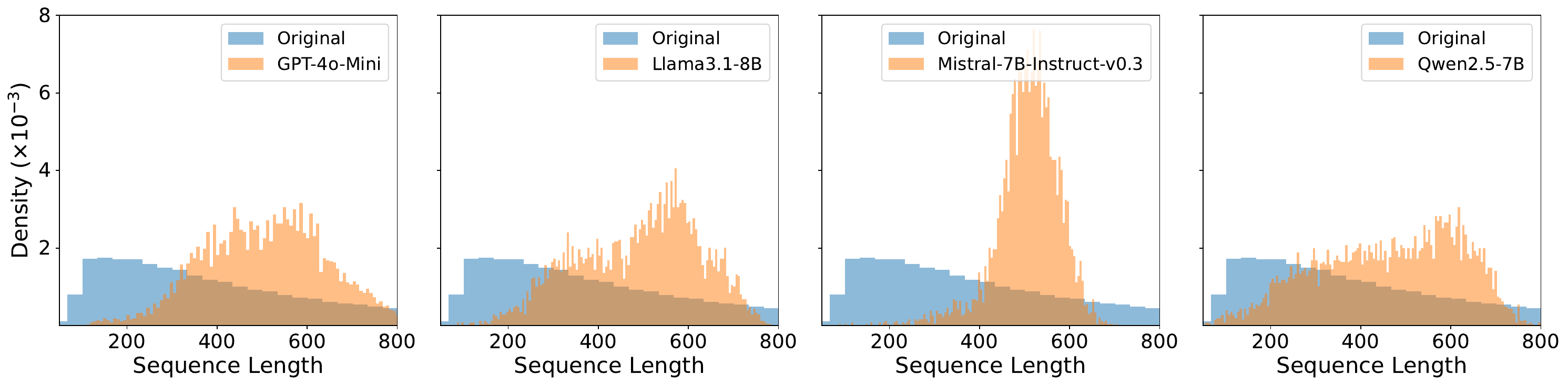}
    \caption{Synthetic sequence-length distributions across different generator models.}
    \label{fig:seq_len_yelp_models}
\end{figure}

Figure~\ref{fig:seq_len_yelp_models} shows the synthetic sequence-length distributions on Yelp for the non-private $\text{SecPE}_{600}$ across different generator models. Among them, \text{Mistral-7B-Instruct-v0.3} exhibits the largest deviation from the original distribution, which aligns with its inferior performance reported in \cref{tab:APIs}.

\section{Additional Theoretical Analysis}
\label{sec:add thms}
\subsection{Further Properties and Relation to DP}
\begin{theorem}[Naive Composition]
Suppose $\mathcal{A}_1: \mathcal{D} \mapsto \mathcal{R}_1$ satisfy $(\vp_1, \vr_1)$-secret protection, $\mathcal{A}_2: \mathcal{D} \times \mathcal{R}_1 \mapsto \mathcal{R}_2$ satisfy $(\vp_2, \vr_2)$-secret protection. Define $\mathcal{A}: \mathcal{D} \to \mathcal{R}_1 \times \mathcal{R}_2$ by 
$$\mathcal{A}(D) = (\mathcal{A}_1(D), \mathcal{A}_2(D, \mathcal{A}_1(D)))$$
Then $\mathcal{A}$ satisfies \((\vp,\vr)\) secret protection such that, coordinate-wise,
\[\vp = \max(\vp_1,\ \vp_2), \quad \vr= \vr_1 + \vr_2.\]
\end{theorem}
\begin{proof} From definition, fix any secret $s_j$, to ensure that both mechanisms \(\mathcal{A}_1\) and \(\mathcal{A}_2\) satisfy their respective $(\vp_1, \vr_1)$ and $(\vp_2, \vr_2)$-secret protection guarantees, the prior distribution \(\pi\) over $\{D_1, \dots, D_K\}$ must satisfy:
\[
\pi(i) \leq p_{1,j}, \quad \text{and} \quad \pi(i) \leq p_{2,j},\quad \forall i\in [K]
\]
Therefore, to apply both guarantees simultaneously, we define the composed mechanism's prior bound as:
\[
p_j = \max(p_{1,j},\ p_{2,j}),
\]
so that any prior distribution satisfying \( \pi(i) \leq p_j \) is valid for both mechanisms.

\noindent Let $B: \mathcal{R}_1 \times \mathcal{R}_2 \to [K]$ be any adversary attempting to identify the index $i$ from the output of $\mathcal{A}(D_i)$. 
Define an intermediate adversary $B_1(y_1) = \arg\max_i \Pr(\mathcal{A}_1(D_i)=y_1)$ (a hypothetical adversary trying to recover $i$ from $y_1$ only). By the assumption that $\mathcal{A}_1$ satisfies $(\vp_1, \vr_1)$-secret protection,
\[ \Pr_{i \sim \pi,\ \mathcal{A}_1}[B_1(\mathcal{A}_1(D_i)) = i] \leq r_{1,j}\]
Since $\mathcal{A}_2$ satisfies $(\vp_2,\vr_2)$-secret protection, for each fixed $y_1$, the success probability of identifying $i$ from $\mathcal{A}_2(D_i, y_1)$ is at most $r_{2,j}$. Hence:
\[\Pr_{i \sim \pi,\ \mathcal{A}_2}[B( \mathcal{A}_2(D_i, y_1), y_1) = i \mid B_1(y_1) \ne i] \leq r_{2,j}\]
Thus, the total success probability of any adversary $B$ satisfies:
\begin{align*}
\Pr_{i \sim \pi,\ \mathcal{A}}[B(\mathcal{A}(D_i)) = i] 
& \leq \Pr_{i \sim \pi,\ \mathcal{A}_1}[B_1(\mathcal{A}_1(D_i)) = i] + \Pr_{i \sim \pi,\ \mathcal{A}_2}[B( \mathcal{A}_2(D_i, y_1), y_1) = i \mid B_1(y_1) \ne i]  \\
& \leq r_{1,j} + r_{2,j}
\end{align*}
Hence, the composed mechanism $\mathcal{A}$ satisfies \((\vp, \vr)\) secret protection such that, coordinate-wise,
\[\vp = \max(\vp_1,\ \vp_2), \quad \vr= \vr_1 + \vr_2.\]
\end{proof}

\begin{remark}
For every $\mu>0$ the following equivalence holds \cite{dong2019gaussiandifferentialprivacy}
\[
\text{the mechanism is }\mu\text{-GDP}
\;\Longleftrightarrow\; (\varepsilon,\delta(\varepsilon))\text{-DP for all }\varepsilon>0,
\]
where
\[
\delta(\varepsilon)
  \;=\;
  \Phi\!\Bigl(-\frac{\varepsilon}{\mu}+\frac{\mu}{2}\Bigr)
  \;-\;
  e^{\varepsilon}\,
  \Phi\!\Bigl(-\frac{\varepsilon}{\mu}-\frac{\mu}{2}\Bigr).
\]
Setting $\mu = \Phi^{-1}(1-p_j) - \Phi^{-1}(1-r_j)$ links the $(\epsilon, \delta)$-DP to \((p_j,r_j)\)-secret protection.
\end{remark}

Consider the same setting as Lemma~\ref{lem:mugdp_to_secdp}: the neighboring datasets $D_1 \simeq D_2$ differ in exactly one element, specifically, $s_j \in x_1 \in D_1$ while $s_j \notin x_2 \in D_2$. The following lemma establishes a direct implication from $(\epsilon,\delta)$-DP to $(\vp,\vr)$-secret protection.
\begin{lemma}
    Any $(\epsilon, \delta)$-DP mechanism $\mathcal{A}$ provides at least $(\vp, \vr)$-secret protection, where
    $$r_j = \frac{1}{1+ \left(e^{\varepsilon}+1/c\right)^{-1} \cdot \frac{1-p_j}{p_j}} + c\cdot \delta,\quad \forall c\geq 1$$
\end{lemma}

\begin{proof}
For any secret $s_j$, denote output distributions as:
\[
P_1 := \mathcal{A}(D_1),\quad P_2 := \mathcal{A}(D_2)
\]
and define the prior $\Pr[D_{\text{train}}= D_1] = p_j$, $\Pr[D_{\text{train}}= D_2] = 1 - p_j$.
Let $y = \mathcal{A}(D_{\text{train}})$ denote an output from $\mathcal{A}$. By Bayes' rule, the posterior odds are:
\[
\frac{\Pr[D_{\text{train}}=  D_1 \mid y]}{\Pr[D_{\text{train}}= D_2 \mid y]} =\frac{\Pr(y \mid D_{\text{train}}= D_1)}{\Pr(y \mid D_{\text{train}}= D_2)} \cdot \frac{\Pr(D_{\text{train}}= D_1)}{\Pr(D_{\text{train}}= D2j)} =  \frac{P_1(y)}{P_2(y)} \cdot \frac{p_j}{1 - p_j}
\]
Now partition the output space $\mathcal{Y}$ into two parts:
\[
G := \left\{y \in \mathcal{Y} :\ \left\vert \log \frac{P_1(y)}{P_2(y)} \right\vert \le \epsilon + t\right\},\qquad U := \mathcal{Y} \setminus G
\]
By the definition of $(\varepsilon,\delta)$-DP, we have:
\[
e^{\varepsilon + t} P_2(U) < P_1(U) \le e^\epsilon P_2(U) + \delta, \quad e^{\epsilon + t}P_1(U) < P_2(U) \le e^\epsilon P_1(U) + \delta 
\]
which implies $P_1(U) \leq \delta/\left(e^{\varepsilon+t} - e^{\varepsilon}\right),\ P_2(U) \leq \delta/\left(e^{\varepsilon+t} - e^{\varepsilon}\right)$. For $c\geq 1$, let $t = \ln (1+e^{-\varepsilon-\ln c})$ so that $P_1(U),\ P_2(U) \leq c\delta $. On the good region $G$, we have:
\[
\frac{\Pr[D_{\text{train}}= D_1 \mid y]}{\Pr[D_{\text{train}}= D_2 \mid y]} \le  \left(e^{\varepsilon}+\frac{1}{c}\right) \cdot \frac{p_j}{1 - p_j}
\Rightarrow \Pr[D_{\text{train}}= D_1 \mid y] \le r^\star := \frac{1}{1 +\left(e^{\varepsilon}+1/c\right)^{-1} \cdot \frac{1 - p_j}{p_j}}
\]
Let $\mu$ denote the output distribution of $\mathcal{A}(I_{\text{train}})$,
\[
\Pr[B(\mathcal{A}(D_i)) = i] \leq \int_{\mathcal{Y}} \max\left\{ \Pr[D_{\text{train}} = D_1 \mid y],\ \Pr[D_{\text{train}} = D_2 \mid y] \right\} d\mu(y)
\]
We upper bound this by splitting the integral over $G$ and $U$:
\begin{align*}
\int_{\mathcal{Y}} \max(\cdot)\, d\mu 
&= \int_G \max(\cdot)\, d\mu + \int_U \max(\cdot)\, d\mu \\
& \le r^\star \cdot \mu(G) + 1 \cdot \mu(U)
\le r^\star + c\cdot \delta
\end{align*}
Thus, $\mathcal{A}$ provides $(p_j,\ r_j)$-secret protection in expectation, where
\[
r_j = \frac{1}{1 + \left(e^{\varepsilon}+1/c\right)^{-1} \cdot \frac{1 - p_j}{p_j}} + c\cdot \delta, \quad \forall c\geq 1.
\]
For $\delta=0$, letting $c \to \infty$, $(\epsilon,0)$-DP implies $(\vp,\vr)$-secret protection with
\[
r_j \;=\; \frac{1}{1 + e^{-\epsilon}\,\frac{1-p_j}{p_j}}.
\]

\end{proof}

\subsection{Convergence Analysis}
\label{sec:convergence analysis}
\begin{definition}[Per-round mis-selection rate]\label{def:rho_t}
The per-round mis-selection rate is the worst-case (over private points) probability that the selection event fails at round $t$:
\[
\rho_t \;\triangleq\; \sup_{x\in D_{\text{priv}}}\ \Pr\!\big(\neg\,\mathrm{Sel}_t(x)\,\big|\,S_t\big)\ \in[0,1].
\]
Here the probability is over the algorithmic randomness at round $t$ given $S_t$.
\end{definition}

\begin{claim}
\label{claim1}
Fix a point $x\in D_{\mathrm{pri}}$ and some iteration $t$. Suppose $z^*\in S_t$ is the closest point to $x$, $V=\text{VARIATION}(z^*,L) \cup \{z^*\}$. If $\Vert x-z^*\Vert \ge \eta$, then with probability at least $(1-\rho_t)/2$, some point in $V$ will get noticeably closer to $x$ than $z^*$, i.e.,
$$\min_{z\in V}\Vert x-z\Vert _2 \le \left(1-\frac{\log L}{4d}\right)\Vert x-z^*\Vert_2.$$
\end{claim}

\begin{theorem}
\label{thm:convergence}
Assume that $\log L \ll d$. With probability $\ge 1-\tau$, the non-private cluster-evolution algorithm outputs $S_{\text{syn}}$ with Wasserstein distance $W_p(D_{\mathrm{pri}},S_{\text{syn}})\le \eta$ after $T$ iterations. $\forall p\in[1,\infty]$ whenever 
	\begin{align}
		T\gg \frac{4}{\,1-\max_t\rho_t\,}\cdot\frac{d \log(D/\eta)}{\log L}+\log(N_{\text{priv}}/\tau)\quad (\text{or more generally,}\ \sum_{t=1}^T (1-\rho_t)\ \gg\ \tfrac{4d\log(D/\eta)}{\log L}),
	\end{align}
\end{theorem}
\begin{proof}
The proof of \cref{claim1} and \cref{thm:convergence} follows directly from Theorem 1 in \cite{lin2025differentiallyprivatesyntheticdata}. In the idealized (no mis-selection) case, $T \gg4 \frac{d \log(D/\eta)}{\log L}+\log(N_{\text{priv}}/\tau)$. Accounting for per-round mis-selections at rate $\rho_t$ weakens the contraction by a factor $(1-\rho_t)$, which replaces $T$ with $\sum_{t=1}^{T}(1-\rho_t)$, or in the worst case, scales by $1/(1-\max_t \rho_t)$. 
\end{proof}

\begin{theorem}[Secret Clustering]
Let $\{C_k\}_{k=1}^{K}\triangleq\{(\ve_k,n_k)\}_{k=1}^{K}$ denote the set of public cluster centers with corresponding cluster sizes. Every private vector is clipped as $\hat{\ve}_{\mathrm{pri},i} = \text{Clip}_R(\ve_{\mathrm{pri},i}) = \ve_{\mathrm{pri},i} \cdot \min\{1,R/\Vert\ve_{\mathrm{pri},i} \Vert\}$, and then assigned to its nearest anchor; Let $m_k$ denote the number of private points assigned to anchor $\ve_k$. For every cluster $k$, we release the perturbed statistics:
\begin{equation}
\begin{aligned}
    \tilde{\ve}_k &:= \frac{\,n_k \cdot \ve_k + \sum_{i\in C_k}\hat \ve_{\mathrm{pri},i}}{n_k+m_k} + \xi_k,
\quad
\xi_k\sim\tfrac{2R}{n_k}\cdot\mathcal{N}(0,\sigma^{2}I_d),\\
\tilde n_k  &= n_k + m_k + \eta_k,
\qquad
\eta_k\sim\mathcal{N}(0,\sigma^2).
\end{aligned}
\end{equation}
where $\sigma$ is chosen by Algorithm~\ref{algo: secret noise} with $T=1$. Then Algorithm~\ref{algo:secret-clustering} satisfies $(\vp,\vr)$-secret protection.
\begin{proof}
Let $D_{\mathrm{sec}}$ denote the dataset containing the secret associated with secret $s$, and let $D_{\mathrm{pri}}^\prime := D_{\mathrm{pri}} \setminus D_{\mathrm{sec}}$, where
    \[
    D_{\mathrm{sec}} := \{x \in D_{\mathrm{pri}} \mid s \in x\}.
    \]
For any cluster $k$, suppose that $m_s$ is the number of private data containing $s$ that is assigned to cluster $k$, then the distribution of $m_s$ is $P:=\mathcal{N}(0, \sigma^2)$ for $D_{\mathrm{pri}}^\prime$ and $Q:=\mathcal{N}(\sum_{x_i \in D_{\mathrm{sec}}}\mathbf{1}\{\arg\min_{j}d(\hat{e}_i,e_j) = k\}, \sigma^2)$ for $D_{\mathrm{pri}}$. 
where 
\[
\left\vert\sum_{x_i \in D_{\mathrm{sec}}}\mathbf{1}\{\arg\min_{j}d(\hat{e}_i,e_j) = k\}\right\vert \leq \vert D_{\mathrm{sec}} \vert,
\]
and $\vert D_{\mathrm{sec}} \vert$ is distributed according to $\mu:=\sum_{x_i \in D_{\mathrm{sec}}} \mathrm{Bern}(\rho_i)$. 
Hence, invoking Lemma 4.5 of \cite{choquettechoo2024privacyamplificationmatrixmechanisms}, \(\mathcal{N}\left(\mu,\sigma^2 \right)\) and \(\mathcal{N}(0,\sigma^2) \) forms a dominating pair that bounds the blow-up function for $P$ and $Q$. Similarly, for cluster center:
\begin{align*}
    \left\Vert\frac{\,n_k \cdot e_k + \sum_{i\in C_k}\hat e_{\mathrm{pri},i}}{n_k+m_k} - e_k \right\Vert & = \left\Vert \frac{\sum_{i\in C_k}\hat e_{\mathrm{pri},i} - m_k \cdot e_k}{n_k+m_k} \right\Vert= \left\Vert \frac{\sum_{i\in C_k} \left(\hat e_{\mathrm{pri},i} - e_k\right)} {n_k+m_k}\right\Vert \\
    & \leq \frac{2 R}{n_k+m_k}\cdot \mu   \leq \frac{2 R}{n_k}\cdot \mu 
\end{align*}
For each iteration, since the noise is calibrated to bound the blow-up function, Theorem~2 of \citep{hayes2023boundingtrainingdatareconstruction} directly implies that the mechanism satisfies $(\vp,\vr)$-secret protection.
\end{proof}
\end{theorem}

\end{document}